\documentclass[12pt]{article}
\usepackage{epsfig}
\usepackage{color}
\usepackage{amssymb,amsmath}

\newcommand{\bea}{\begin{eqnarray}}
\newcommand{\eea}{\end{eqnarray}}

 \makeatletter
\def\alt{\mathrel{\mathpalette\gl@align<}}
\def\agt{\mathrel{\mathpalette\gl@align>}}
\def\gl@align#1#2{\lower.6ex\vbox{\baselineskip\z@skip\lineskip\z@
\ialign{$\m@th#1\hfil##\hfil$\crcr#2\crcr\sim\crcr}}} \makeatother

\begin{document}
\begin{flushright}
KEK-TH-1518
\end{flushright}

\vspace*{1.0cm}

\begin{center}
\baselineskip 20pt 
{\Large\bf 
Dark matter in the classically conformal $B-L$ model
}

\vspace{1cm}

{\large 
Nobuchika Okada$^{a,}$\footnote{okadan@ua.edu} 
and 
Yuta Orikasa$^{b,}$\footnote{orikasa@post.kek.jp}
} \vspace{.5cm}

{\baselineskip 20pt \it
$^{a}$ Department of Physics and Astronomy, 
University of Alabama, \\ 
Tuscaloosa,  Alabama 35487, USA \\ 
$^{b}$ KEK Theory Center,  
High Energy Accelerator Research Organization (KEK)  \\
and \\
Department of Particles and Nuclear Physics, \\
The Graduate University for Advanced Studies (SOKENDAI), 
\\
1-1 Oho, Tsukuba, Ibaraki 305-0801, Japan 
}

\vspace{.5cm}

\vspace{1.5cm} {\bf Abstract}
\end{center}

When the classically conformal invariance is imposed 
 on the minimal gauged $B-L$ extended Standard Model (SM), 
 the $B-L$ gauge symmetry is broken 
 by the Coleman-Weinberg mechanism naturally at the TeV scale. 
Introducing a new $Z_2$ parity in the model, 
 we investigate phenomenology of a right-handed neutrino 
 dark matter whose stability is ensured by the parity. 
We find that the relic abundance of the dark matter particle 
 can be consistent with the observations 
 through annihilation processes enhanced by resonances 
 of either the SM Higgs boson, 
 the $B-L$ Higgs boson or the $B-L$ gauge boson (Z' boson). 
Therefore, the dark matter mass is close to half of 
 one of these boson masses. 
Due to the classically conformal invariance and 
 the $B-L$ gauge symmetry breaking via the Coleman-Weinberg mechanism, 
 Higgs boson masses, Z' boson mass and the dark matter mass 
 are all related, and we identify the mass region 
 to be consistent with experimental results. 
We also calculate the spin-independent cross section 
 of the dark matter particle off with nucleon and 
 discuss implications for future direct dark matter 
 search experiments.

\thispagestyle{empty}

\newpage

\addtocounter{page}{-1}
\setcounter{footnote}{0}

\baselineskip 18pt

\section{Introduction}

The minimal gauged $B-L$ extended Standard Model (SM) 
 is one of very attractive scenarios beyond the SM 
 and has been receiving fare amount of attentions these days. 
The model is an elegant and simple extension of the SM, 
 in which the right-handed neutrinos of three generations 
 are necessarily introduced for the cancellation
 of the gauge and gravitational anomalies. 
In addition, the mass of right-handed neutrinos arises 
 associated with the U(1)$_{B-L}$ gauge symmetry breaking 
 and the seesaw mechanism \cite{seesaw} 
 for a natural generation of tiny neutrino masses 
 is automatically implemented. 
The Large Hadron Collider (LHC) is currently exploring 
 the TeV scale physics by collecting data very rapidly. 
In the view point of LHC physics 
 it is the most desirable that the $B-L$ symmetry breaking scale 
 lies at the TeV scale, 
 so that the $B-L$ gauge boson (Z' boson) and the right-handed 
 neutrinos can be discovered in the near future \cite{BL-LHC}.

Recently, the minimal $B-L$ model with the classically conformal 
 invariance has been proposed \cite{IOO1}
 and it has been shown \cite{IOO2} that the $B-L$ symmetry breaking 
 in the model is naturally realized at the TeV scale 
 when the $B-L$ gauge coupling constant is of the same order of 
 the SM gauge coupling constants. 
Furthermore, one of cosmological aspects of the minimal $B-L$ model 
 at the TeV scale, baryogenesis via leptogenesis, 
 has been investigated with detailed numerical analysis. 
It has been shown \cite{IOO3} that the observed baryon 
 asymmetry in the present universe can be reproduced 
 via the resonant leptogenesis \cite{RLG} 
 with a suitable set of model parameters, 
 which  is also consistent with the neutrino oscillation data.

Towards the completion of phenomenology for 
 the $B-L$ model at the TeV scale, 
 we investigate, in this paper, another cosmological aspect, 
 namely the dark matter issue. 
Among many possibilities, a very concise way to introduce 
 a dark matter candidate in the minimal $B-L$ model 
 has been proposed in Ref.~\cite{OS}, 
 where a new $Z_2$ parity, 
 instead of new particle(s) for the dark matter candidate, 
 is introduced. 
Under the the parity, one right-handed neutrino is assigned 
 as odd while all other particles are even. 
This parity assignment makes the $Z_2$-odd right-handed neutrino 
 stable and hence the candidate for the (cold) dark matter. 
It has been found \cite{OS} that the observed relic abundance 
 of the right-handed neutrino dark matter can be achieved 
 through interactions with Higgs bosons. 
In this paper, we adopt this idea to the classically conformal 
 $B-L$ extended model and investigate phenomenology of 
 the right-handed neutrino dark matter. 
Although our analysis is quit analogous to 
 those in Ref.~\cite{OS}, the classically conformal invariance 
 imposed on the model plays the crucial role 
 to severely constrain the parameter space of the model.

The paper is organized as follows. 
In the next section, we give a brief review on 
 the classically conformal $B-L$ extended model 
 (with $Z_2$ parity) and the natural realization 
 of the $B-L$ symmetry breaking at the TeV scale. 
In Sec.~3, we analyze the relic abundance of 
 the right-handed neutrino  dark matter and identify 
 the parameter region for reproducing 
 the observed dark matter abundance. 
We also calculate the spin-independent scattering cross section 
 between the dark matter particle and nucleon in Sec~4. 
The last section is devoted for summary.

\section{The classically conformal $B-L$ extended SM}

The minimal $B-L$ extended SM is based on the gauge group 
 SU(3)$_c\times$SU(2)$_L \times $U(1)$_Y \times$U(1)$_{B-L}$. 
As has been discussed above, we introduce a global $Z_2$ parity 
 in the model, and the particle contents are listed in Table~1. 
The three right-handed neutrinos ($\nu_R^k (k=1,2,3)$) 
 are necessarily introduced to cancel all the gauge and 
 gravitational anomalies. 
Only the $\nu_R^1$ is assigned to be odd under the $Z_2$ parity. 
The SM singlet scalar field ($\Phi$) works to break 
 the U(1)$_{B-L}$ gauge symmetry by its vacuum expectation value (VEV), 
 $\langle \Phi \rangle =v_{B-L}/\sqrt{2}$. 
Once the $B-L$ gauge symmetry is broken, 
 the Z' boson acquires mass, 
\bea
  m_{Z'} = 2 g_{B-L} v_{B-L},  
\label{Z'mass}
\eea
 where $g_{B-L}$ is the $B-L$ gauge coupling constant. 
The LEP experiment has set the lower bound 
 on the $B-L$ symmetry breaking scale as 
 $v_{B-L} \gtrsim$ 3 TeV \cite{vBL}. 
Recent LHC results for Z' boson search 
 with 1.1 fb$^{-1}$ \cite{LHCZ'} 
 excluded the $B-L$ Z' gauge boson mass 
 $m_{Z'} \lesssim 1.5$ TeV \cite{Basso} 
 when the $B-L$ coupling is not too small. 
We see that the LEP bound is more severe 
 than the LHC bound for $m_{Z'} \gtrsim 1.5$ TeV. 

\begin{table}[t]
\begin{center}
\begin{tabular}{c|ccc|c|c}
  & SU(3)$_c$ & SU(2)$_L$ & U(1)$_Y$ & U(1)$_{B-L}$ & $Z_2$  \\
\hline
$ q_L^i $    & {\bf 3}   & {\bf 2}& $+1/6$ & $+1/3$ &+ \\ 
$ u_R^i $    & {\bf 3} & {\bf 1}& $+2/3$ & $+1/3$  & + \\ 
$ d_R^i $    & {\bf 3} & {\bf 1}& $-1/3$ & $+1/3$  & + \\ 
\hline
$ \ell^i_L$    & {\bf 1} & {\bf 2}& $-1/2$ & $-1$  & + \\ 
$ \nu_R^1 $   & {\bf 1} & {\bf 1}& $ 0$   & $-1$   & $-$ \\ 
$ \nu_R^j $   & {\bf 1} & {\bf 1}& $ 0$   & $-1$   & + \\ 
$ e_R^i   $   & {\bf 1} & {\bf 1}& $-1$   & $-1$   & + \\ 
\hline 
$ H$         & {\bf 1} & {\bf 2}& $+1/2$  &  $ 0$ & + \\ 
$ \Phi$      & {\bf 1} & {\bf 1}& $  0$  &  $+2$  & + \\ 
\end{tabular}
\end{center}
\caption{
Particle contents:  
In addition to the SM particles, 
 three right-handed neutrinos, $\nu_R^1$ and $\nu_R^j$ ($j=2,3$),  
 and a complex scalar $\Phi$ are introduced. 
Under the global $Z_2$ parity, $\nu_R^1$ is assigned to be odd, 
 while the other particles are even. 
$i=1,2,3$ is the generation index. 
}
\end{table}

The Lagrangian relevant for the seesaw mechanism is given by
\bea 
  {\cal L} \supset -y_D^{ij} \overline{\nu_R^i} H \ell_L^j  
 - \frac{1}{2} y_N^k \Phi \overline{\nu_R^{k c}} \nu_R^k 
 +{\rm h.c.},  
\label{Yukawa}
\eea 
 where without loss of generality, we work on the basis 
 in which the second term is diagonalized and $y_N^k (k=1,2,3)$ 
 is real and positive. 
The first term gives the neutrino Dirac mass matrix  
 after the electroweak symmetry breaking. 
Note that because of $Z_2$ parity, $\nu_R^1$ has 
 no coupling with the lepton doublets 
 and the neutrino Dirac mass matrix is 2 by 3. 
The right-handed neutrino Majorana masses are generated 
 through the second term associated 
 with the $B-L$ gauge symmetry breaking 
 ($m_{N_i} = \frac{y_N^i}{\sqrt{2}} v_{B-L}$).

The $B-L$ symmetry breaking scale is determined by parameters 
 in the (effective) Higgs potential and in general 
 we can take any scale for it as long as the experimental
 constraints are satisfied. 
It has been pointed out in \cite{IOO1, IOO2}
 if we impose the classically conformal symmetry 
 on the minimal $B-L$ model, the $B-L$ symmetry breaking 
 is naturally realized at the TeV scale. 
Thus, the mass scale of all new particles are  
 at the TeV scale or smaller.

Under the hypothesis of the classically conformal invariance 
 of the model, the classical scalar potential is given by 
\begin{equation}
 V(H,\Phi)=\lambda_H \left(H^\dagger H\right)^2
  +\lambda \left(\Phi^\dagger\Phi\right)^2
  -\lambda^\prime \left(\Phi^\dagger\Phi\right)\left(H^\dagger H\right).
\end{equation}
Since there is no mass term in the Higgs potential, 
 the symmetry should be broken radiatively through 
 the Coleman-Weinberg (CW) mechanism \cite{CW}. 
Assuming a small $\lambda^\prime$, 
 in which case the SM Higgs sector and the $B-L$ Higgs 
 sector are approximately decoupled, 
 the renormalization group (RG) improved effective potential 
 for the $B-L$ sector gives the stationary condition \cite{IOO1}, 
\begin{eqnarray}
 \alpha_\lambda \simeq-\frac{6}{\pi} 
 \left(\alpha_{B-L}^2-\frac{1}{96}\sum_i\left(\alpha^i_N\right)^2
 \right), 
\label{coupling relation}
\end{eqnarray}
 where $\alpha_\lambda=\lambda/(4\pi)$, $\alpha_{B-L}=g_{B-L}^2/(4\pi)$ 
 and $\alpha_N^i=\left(y_N^i\right)^2/(4\pi)$ 
 are the RG running coupling evaluated at $v_{B-L}$. 
The mass of the SM singlet Higgs is given by 
\begin{equation}
 m_\phi^2 
\simeq  
-16\pi\alpha_\lambda v_{B-L}^2 
 \simeq \frac{6}{\pi} 
 \left( \alpha_{B-L} - \frac{1}{96} 
 \frac{\sum_i(\alpha_N^i)^2}{\alpha_{B-L}}
\right) m_{Z'}^2 
 \simeq \frac{6}{\pi} \alpha_{B-L} m_{Z'}^2 
\label{Phimass}
\end{equation} 
Note that the $B-L$ symmetry breaking via the CW mechanism 
 leads to the mass relation between the SM singlet 
 Higgs and Z' boson.

Once $v_{B-L}$ is generated, the SM Higgs doublet 
 acquires a mass squared, $-\lambda^\prime v_{B-L}^2$, 
 so that the electroweak symmetry is broken 
 for $\lambda^\prime > 0$. 
The SM Higgs boson mass is given by 
\begin{equation}
  m_h^2= \lambda^\prime v_{B-L}^2=2\lambda_H v^2, 
\end{equation}
 with $v=246$ GeV, 
 and the scalar mass matrix is found to be 
\begin{equation}
 {\cal M}=
\left(
\begin{array}{cc}
 m_h^2 & -m_h^2 \left( \frac{v}{v_{B-L}} \right) \\
 -m_h^2 \left( \frac{v}{v_{B-L}} \right) & m_\phi^2 
\end{array}\right).  
\end{equation}
Thus the mixing angle to diagonalize the mass matrix is given by 
\begin{equation}
 \tan 2 \theta 
 =\frac{2m_h^2(v/v_{B-L})}{m_h^2-96\alpha_{B-L}^2v_{B-L}^2}. 
\end{equation}
Using Eqs.(\ref{Z'mass}) and (\ref{Phimass}), 
 the mass matrix and the mixing angle are functions of 
 three independent parameters, $m_h$, $\alpha_{B-L}$ and $m_{Z'}$. 
Except for the special case, 
 $ m_h^2 \simeq 96 \alpha_{B-L}^2v_{B-L}^2$, 
 the mixing angle is always small because of the suppression 
 by $v/v_{B-L}$ with $v=246$ GeV and $v_{B-L} \gtrsim 3$ TeV. 
Thus, one mass eigenstate is the SM-like Higgs boson 
 while the other is the SM singlet-like Higgs boson.

\begin{figure}[t]\begin{center}
\includegraphics[scale=0.6]{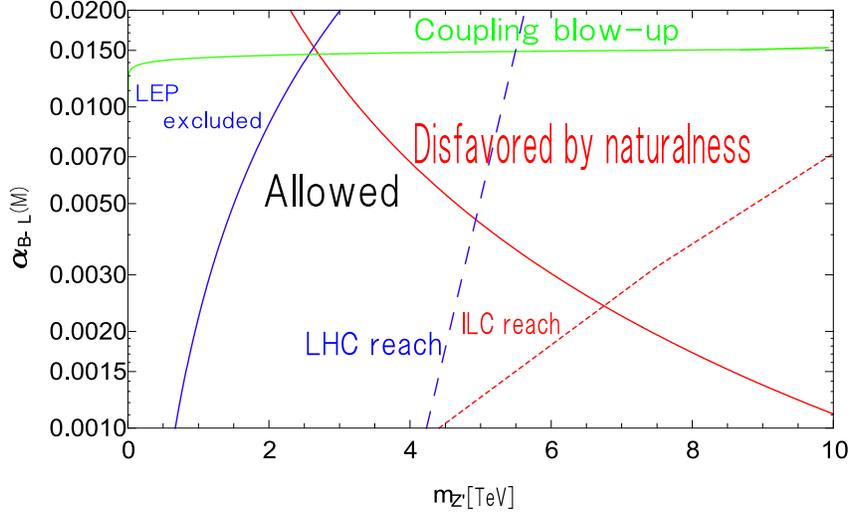}
\caption{
The allowed parameter region is drawn \cite{IOO2}. 
The upper region of the almost straight line (in green) 
 is rejected by a requirement that the $B-L$ gauge coupling 
 does not diverge up to the Planck scale.
The upper-right side of the solid line (in red)
 is disfavored by the naturalness condition of the electroweak scale.
The left of the solid line (in blue) has been already excluded
 by the LEP experiment, $v_{B-L} \gtrsim 3$ TeV. 
Recent LHC results for Z' boson search excluded the region 
 $m_{Z'} \lesssim 1.5$ TeV \cite{Basso}. 
The left of the dashed line can be explored in 5-$\sigma$ significance 
 at the LHC with $\sqrt{s}$=14 TeV and an integrated 
 luminosity 100 fb$^{-1}$.  
The left of the dotted line can be explored at the ILC 
 with $\sqrt{s}$=1 TeV, assuming 1\% accuracy. 
}
\label{AllowedRegion}
\end{center}
\end{figure}

There are theoretical constraints on $\alpha_{B-L}$ and $m_{Z'}$. 
First we require that the $B-L$ gauge coupling does not 
 blow up below the Planck scale ($M_{Pl}$). 
Second, the Higgs boson mass can receive 
 big quantum corrections at two loop diagrams 
 involving top quark and Z' boson such as \cite{IOO1}
\begin{eqnarray}
 \Delta m_h^2 =
 \frac{8\alpha_{B-L}m_t^2 m_{Z^\prime}^2}
 {\left( 4\pi\right)^3 v^2}
 \log\frac{M_{Pl}^2}{m_{Z^\prime}^2}. 
\label{2-loop}
\end{eqnarray} 
Here we have used the Planck scale for the cutoff of the loop integral. 
In the naturalness point of view, this corrections 
 should not exceed the electroweak scale 
 and we obtain a stringent bound on Z' boson mass 
 by imposing $ \Delta m_h^2 \leq v/\sqrt{2}$, for example. 
The allowed parameter region is depicted in Fig.~\ref{AllowedRegion}. 
The upper region $\alpha \gtrsim 0.015$ is excluded 
 by the condition of the $B-L$ gauge coupling blow-up. 
The left of the solid line (in blue) is excluded by 
 the LEP experiment, $v_{B-L} \gtrsim 3$ TeV. 
The upper-right side of the solid line (in red) 
 is disfavored by the naturalness condition 
 of the electroweak scale. 
The future search reach of the Z' boson mass 
 is also shown on the figure. 
The left of the dashed line can be explored 
 in 5-$\sigma$ significance at the LHC with $\sqrt{s}$=14 TeV 
 and an integrated luminosity 100 fb$^{-1}$ \cite{LHCreach}.  
The left of the dotted line can be explored 
 at the International Linear Collider with $\sqrt{s}$=1 TeV, 
 assuming 1\% accuracy \cite{LHCreach}.  
The figure indicates that if the $B-L$ gauge coupling 
 is of the same order as the SM gauge couplings, 
 $Z^\prime$ boson mass appears around a few TeV. 

\section{Relic density of right-handed neutrino dark matter}

The $Z_2$-odd right-handed neutrino is stable 
 and the dark matter candidate. 
In this section, we estimate its relic abundance 
 and identify the model parameters to be consistent 
 with the current observations. 
The dark matter particles annihilate into the SM particles 
 through interactions with the Z' boson and the Higgs bosons. 
In practice, the annihilation processes are dominated by 
 the $s$-channel mediated by the $B-L$ gauge boson 
 and the Higgs bosons. 
All the general formulas of the annihilation cross sections 
 necessary for our analysis are listed 
 in Appendices of Ref.~\cite{OS}.

The Boltzmann equation of the right-handed neutrino dark matter 
 is given by 
\begin{equation}
 \frac{dY_{N_1}}{dz}=-\frac{z\langle\sigma v\rangle s}{H(m_{N_1})}
 \left(Y_{N_1}^2-Y_{N_1}^{eq 2}\right), 
\end{equation}
 where $Y_{N_1}$ is the yield (the ratio of the number density to 
 the entropy density $s$) of the right-handed neutrino dark matter, 
 $Y_{N_1}^{eq}$ is the yield in thermal equilibrium, 
 temperature of the universe is normalized by the mass of the
 right-handed neutrino $z=m_{N_1}/T$, 
 $H(m_{N_1})$ is the Hubble parameter at $T=m_{N_1}$, 
 and $\langle\sigma v\rangle$ is the thermal averaged product 
 of the annihilation cross section and the relative velocity. 
The density parameter of the dark matter particle is written as 
\begin{equation}
 \Omega_{DM}h^2=\frac{m_{N_1} s_0 Y_{N_1}(\infty)}{\rho_c/h^2}, 
\end{equation}
 where $Y_{N_1}(\infty)$ is the asymptotic value of the yield, 
 $s_0=2890$cm$^{-3}$ is the entropy density of the present universe, 
 and $\rho_c/h^2=1.05\times10^{-5}$GeV cm$^{-3}$ is the critical density. 
The thermal relic abundance of the dark matter is approximately given by 
\begin{equation}
 \Omega_{DM}h^2=1.1 
\times10^9\frac{m_{N_1}/T_d}
 {\sqrt{\mathstrut g_*} M_{Pl} \langle\sigma v\rangle},  
\end{equation}
 where $g_*$ is the total number of relativistic degrees of 
 freedom, and $T_d$ is the decoupling temperature. 

\begin{figure}[t]
\begin{tabular}{cc}
\begin{minipage}{0.5\hsize}
\begin{center}
\includegraphics[scale=0.8]{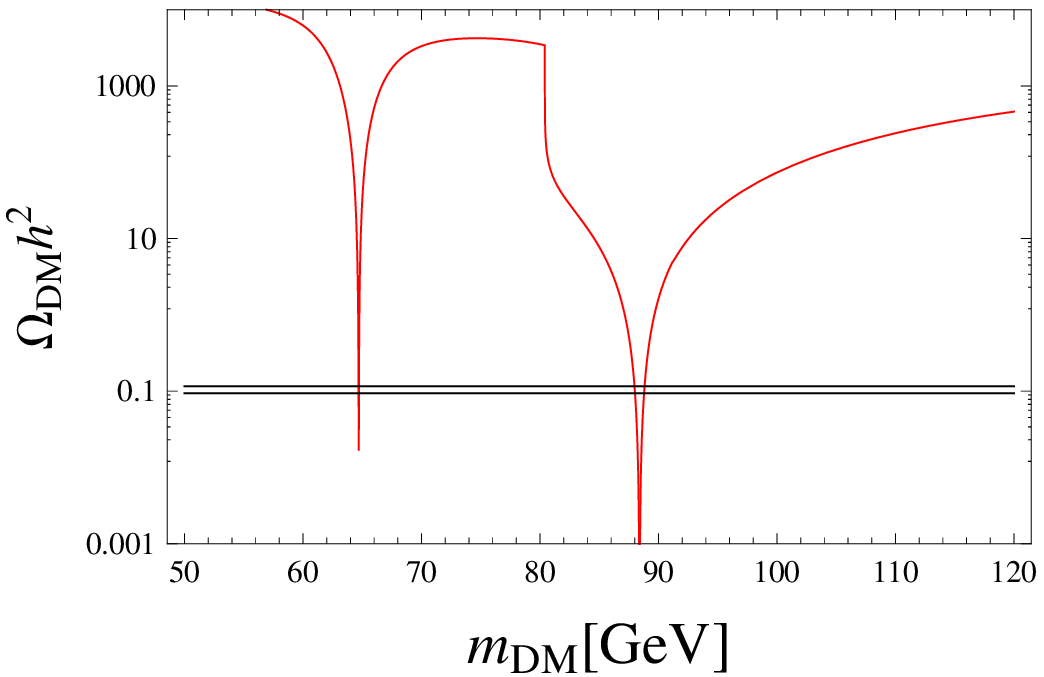}
\end{center}
\end{minipage}
\begin{minipage}{0.5\hsize}
\begin{center}
\includegraphics[scale=0.8]{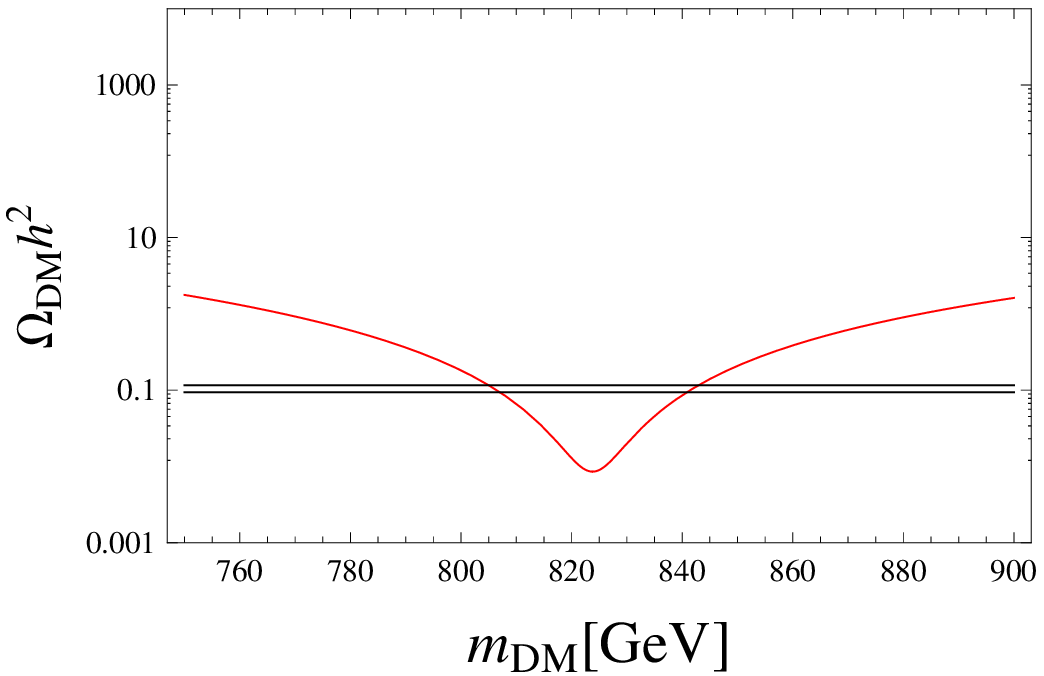}
\end{center}
\end{minipage}
\end{tabular}
\caption{
The thermal relic density of the right-handed neutrino dark matter 
 as a function of its mass. 
The left panel corresponds to the Higgs resonance regions 
 while the right panel to the $Z^\prime$ resonance region. 
}
\label{DMLEP}
\end{figure}

There are six free parameters involved in our analysis: 
 $m_h, \alpha_{B-L}, m_{z^\prime}, m_{N_1}, m_{N_2}, m_{N_3}$. 
For simplicity, we fix three of them as follows:   
\begin{equation}
  m_h=130~{\rm GeV}, \; \; m_{N_2}=m_{N_3}=2~\rm{TeV}. 
\end{equation} 
Note that finely degenerate masses for 
 the two $Z_2$-even right-handed neutrinos are 
 necessary for the successful baryogenesis 
 via resonant leptogenesis \cite{RLG}. 
Then, we have only three free parameters left, 
 namely, $\alpha_{B-L}$, $m_{Z^\prime}$ 
 and $m_{N_1}=m_{DM}$ being the dark matter mass. 
As we have discussed in the previous section, 
 $\alpha_{B-L}$ and $m_{Z^\prime}$ are constrained 
 as depicted in Fig.~\ref{AllowedRegion}. 
In the following analysis, we show our results 
 along three lines in Fig.~\ref{AllowedRegion}: 
 the ``LEP line'' due to the constraint $v_{B-L}=3$, 
 the ``Naturalness line'' and the ``LHC line'' 
 corresponding to the LHC search reach. 
Along these lines, $\alpha_{B-L}$ is given 
 as a function of $m_{Z^\prime}$, 
 so that we show our results in terms of 
 only two free parameters, $m_{Z'}$ and $m_{DM}$.

Fig.~\ref{DMLEP} shows the resultant relic density 
 $\Omega_{DM}h^2$ as a function of the dark matter mass 
 for fixed values of $\alpha_{B-L}=0.006$ and 
 $m_{Z'} =1.65$ TeV (on the LEP line), 
 along with the observed values at 2-$\sigma$ level 
 \cite{WMAP}
\bea 
 \Omega_{DM}^{obs} h^2 = 0.1120 \pm 0.0056.  
\eea
We find that the observed relic abundance can be achieved 
 only if the dark matter mass is very close to 
 the resonance point of the $s$-channel annihilation 
 process mediated by either the SM-like Higgs boson, 
 the SM singlet-like Higgs boson or the Z' boson.

\begin{figure}[t]
\begin{tabular}{cc}
\begin{minipage}{0.5\hsize}
\begin{center}
\includegraphics[scale=.7]{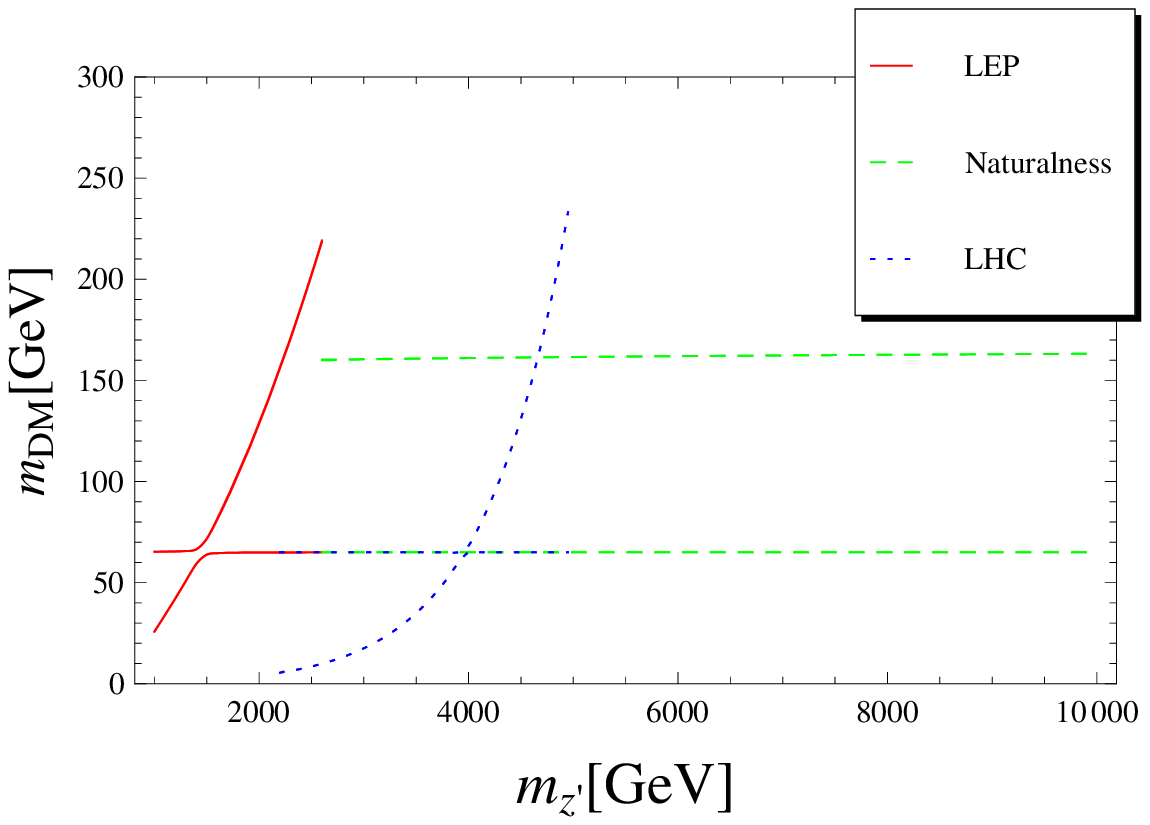}
\end{center}
\end{minipage}
\begin{minipage}{0.5\hsize}
\begin{center}
\includegraphics[scale=.7]{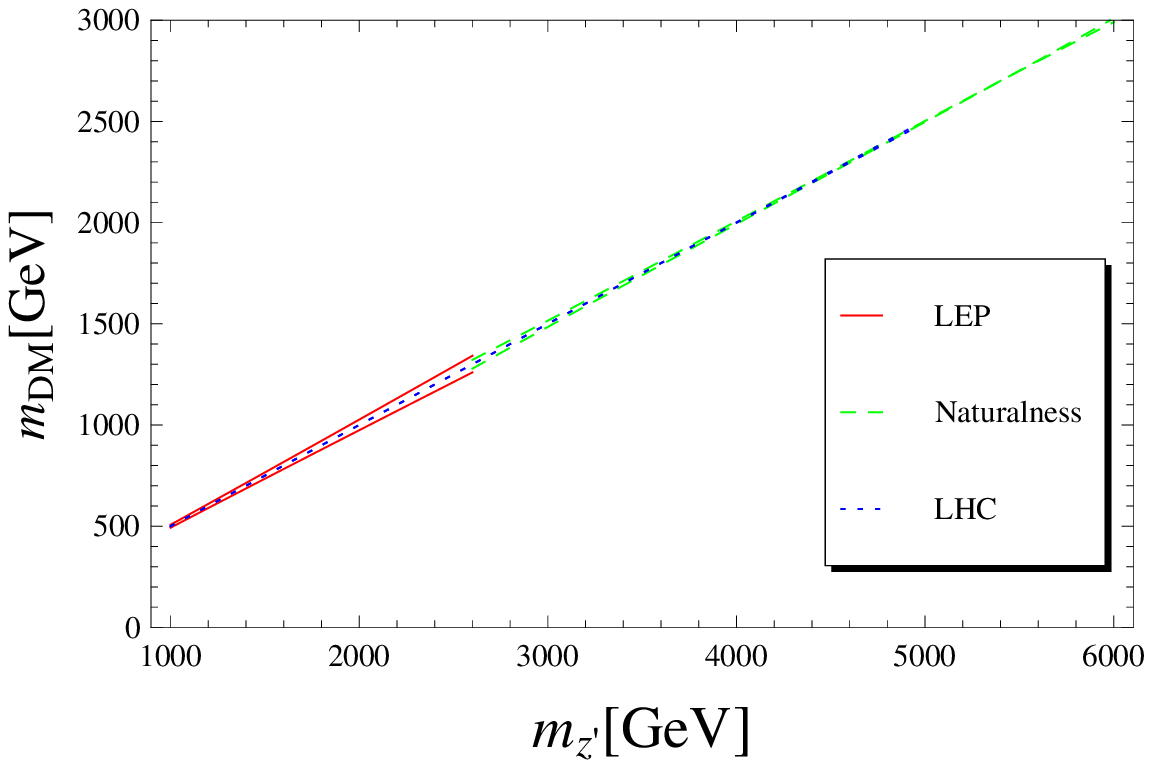}
\end{center}
\end{minipage}
\end{tabular}
\caption{
The dark matter mass as a function of $Z^\prime$ boson mass. 
The left panel shows the results when the dark matter mass 
 is close to half of Higgs masses, 
 while the right panel corresponds to the case 
 with the dark matter mass being close 
 to half of the $Z^\prime$ boson mass. 
The solid, dashed and dotted lines correspond to 
 the LEP, Naturalness and LHC lines, respectively. 
} 
\label{DMmass}
\end{figure}

Along the three lines, we determine the dark matter mass 
 by comparing it with the observed relic abundance. 
The results are shown in Fig.~\ref{DMmass} 
 as a function of the Z' boson mass. 
The left panel shows the solution 
 when the dark matter mass is close to the Higgs 
 resonance points, while to the Z' boson resonance point 
 in the right panel. 
The solid, dashed and dotted lines correspond to 
 the LEP, Naturalness and LHC lines, respectively. 
As can be seen in Fig.~\ref{DMLEP}, there are two solutions 
 corresponding to each the resonance point, 
 but they are well-overlapped and not distinguishable  
 for most of lines in Fig.~\ref{DMmass}. 
Along the Naturalness line in the left panel, 
 Higgs boson masses are almost constant 
 as can be understood from Eqs.~(\ref{Phimass}) and (\ref{2-loop}).

\section{Direct detection of dark matter}

A variety of experiments are underway and also planned 
 to detect a dark matter particle directly or indirectly,
 through the elastic scattering of dark matter particle 
 off with nuclei. 
The right-handed neutrino dark matter in our model 
 couples with quarks in two ways. 
One is through Higgs bosons, the other is via Z' boson exchange. 
Because of its Majorana nature, 
 the dark matter particle has the axial vector coupling 
 with the Z' boson, while the quarks have the vector coupling. 
As a result, there is no contribution from 
 the Z' boson exchange in the non-relativistic limit. 
Therefore, we consider only the spin-independent 
 elastic scattering process via Higgs boson exchange.

The spin-independent dark matter-proton cross section is given by 
\begin{equation}
 \sigma_{SI}^{(p)}=\frac{4}{\pi}\left(\frac{m_pm_N}{m_p+m_N}\right)^2
 f_p^2, 
\end{equation}
with the hadronic matrix element 
\begin{equation}
 \frac{f_p}{m_p}=\sum_{q=u,d,s}f_{Tq}\frac{\alpha_q}{m_q}+\frac{2}
  {27}f_{TG}\sum_{q=c,b,t}\frac{\alpha_q}{m_q}, 
\end{equation}
and the effective vertex 
\bea
 \alpha_q=- \frac{y_N^1 y_q}{4} \sin 2 \theta  
 \left(\frac{1}{M_H^2}-\frac{1}{M_\Phi^2} \right), 
\eea
where $m_q$ is a mass of a quark with a Yukawa coupling $y_q$, 
 and $M_H (M_\Phi)$ is the mass eigenvalue of 
 the SM-like (SM singlet-like) Higgs boson.  
The parameter $f_{Tq}$ has recently been 
 evaluated accurately by the lattice QCD simulation 
 using the overlap fermion formulation. 
The result of the simulation has shown that $f_{Tu}+f_{Td}\simeq0.056$ 
 and $| f_{Ts}| \leq 0.08$ \cite{QCD sim1}. 
On the other hand, the parameter $f_{TG}$ is obtained by $f_{Tq}$ 
 through the trace anomaly, $1=f_{Tu}+f_{Td}+f_{Ts}+f_{TG}$
 \cite{QCD sim2}. 
For conservative analysis, we take $ f_{Ts}=0$. 

\begin{figure}[t]
\begin{center}
\includegraphics[scale=.79]{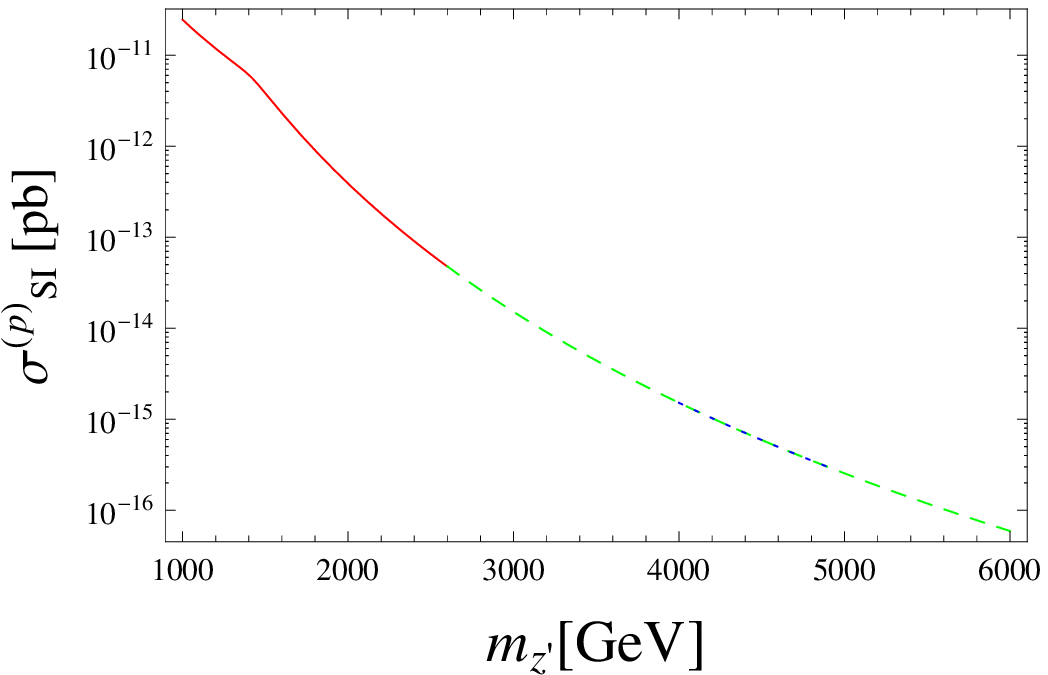}
\includegraphics[scale=.79]{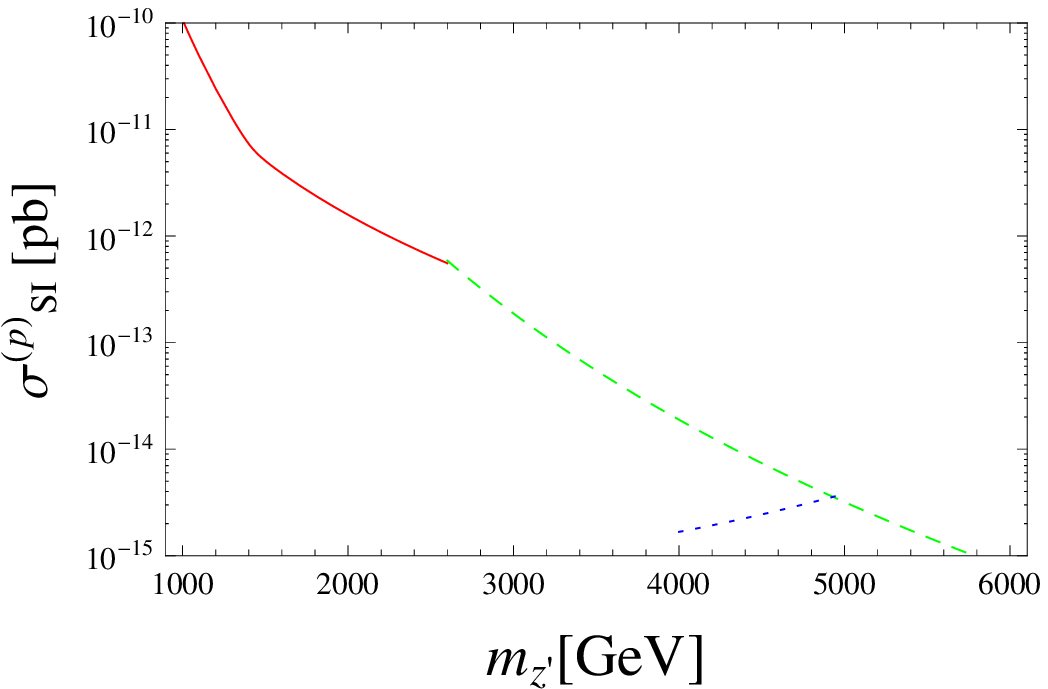}
\includegraphics[scale=.79]{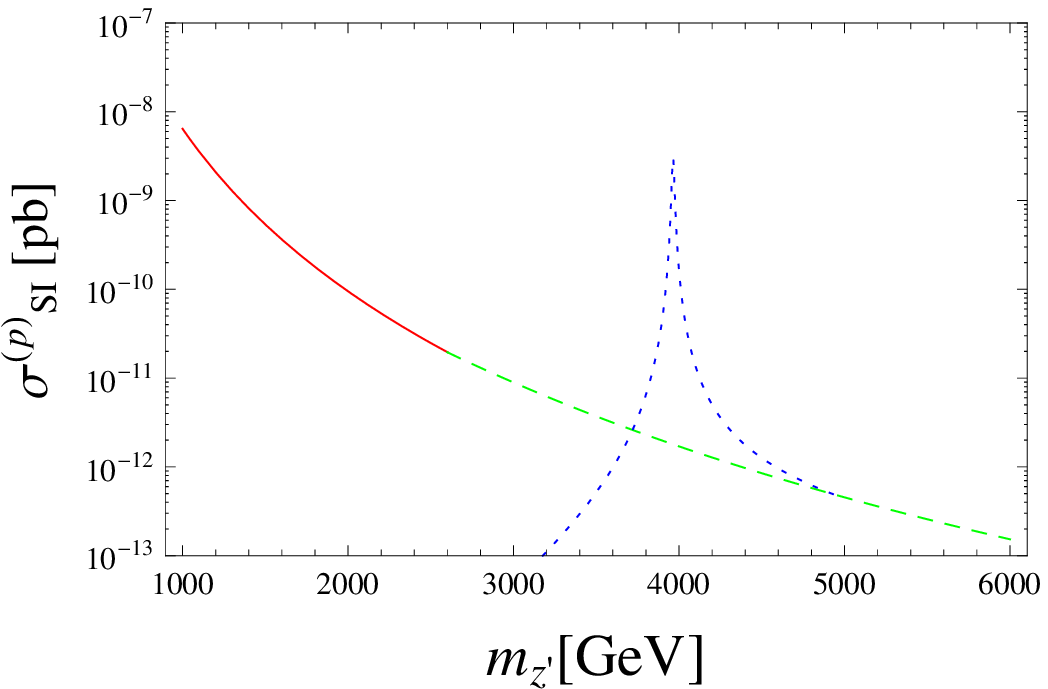}
\end{center}
\caption{
 The spin-independent cross sections as a function 
 of Z' boson mass. 
The top-left panel shows the case of $m_{DM} \sim M_H/2$, 
 the top-right panel is for the case of $M_{DM} \sim M_\Phi/2$, 
 and the case for $M_{DM} \sim m_{Z'}/2$ is shown 
 in the bottom panel.  
The (red) solid lines, the (green) dashed lines 
 and the (blue) dotted lines correspond to 
 the LEP line, Naturalness line and LHC line, 
 respectively. 
}
\label{DMDD}
\end{figure}

The results for the spin-independent cross sections 
 are depicted in Fig.~\ref{DMDD}. 
The top-left (top-right) panel shows the cross sections 
 along the LEP, Naturalness and LHC lines 
 for the case with $m_{DM} \sim M_H/2$ ($m_{DM} \sim M_\Phi/2$). 
The resultant cross section is found to be far below the current limits 
 reported by XENON100 \cite{XENON100}: 
 $\sigma_{SI} \lesssim 10^{-8}-10^{-7}$ pb, 
 for a dark matter mass of 100 GeV$-$1 TeV. 
The result for the case with $m_{DM} \sim m_{Z'}/2$ 
 is depicted in the bottom panel. 
The cross section found in this case is relatively  higher 
 but yet below the current limit. 
The cross section is enhanced around $m_{Z'} =4$ TeV 
 where the mixing angle becomes maximum. 
In future experiments such as XENON1T \cite{XENON1T} 
 the search limit can be as low as $\sigma_{SI} = 10^{-11}-10^{-10}$ pb 
 and the region $m_{Z'} \lesssim 3$ TeV in the bottom panel 
 can be tested.

\section{Summary}

Gauged $B-L$ extension of the Standard Model is 
 a very simple and elegant way 
 to account for the mass and flavor mixing of neutrinos. 
Three right-handed neutrinos are introduced 
 to make the model free from the gauge and gravitational anomalies. 
Associated with the $B-L$ gauge symmetry breaking, 
 the right-handed neutrinos acquire the Majorana mass,  
 and after the electroweak symmetry breaking 
 the light neutrino masses are generated 
 through the seesaw mechanism. 
The scale of the $B-L$ gauge symmetry breaking 
 is arbitrary as long as phenomenological constraints 
 are satisfied.

We have introduced the classically conformal invariance 
 of the model, which forbids the mass terms 
 for the Standard Model Higgs doublets and 
 the SM singlet $B-L$ Higgs field. 
In this system, the $B-L$ gauge symmetry is radiatively 
 broken via the CW mechanism, 
 which triggers the electroweak symmetry breaking 
 by generating the negative mass squared 
 for the SM Higgs doublet. 
The naturalness argument constrains 
 the model parameter space and we find 
 the $B-L$ symmetry breaking scale to be the TeV scale 
 if the $B-L$ gauge coupling is of the same order 
 of the SM gauge couplings. 
Therefore, all new particles in the model, 
 Z' gauge boson, right-handed neutrinos and $B-L$ Higgs boson,
 have masses around TeV or smaller and 
 they can be discovered at the LHC.

We have investigated cosmological aspects of the model, 
 in particular, the dark matter issue. 
We have introduced a $Z_2$ parity 
 under which one right-handed neutrino is odd 
 and the other particles are even. 
In this way, the $Z_2$-odd right-handed  neutrino 
 becomes the candidate for the dark matter, 
 without introducing any new particles into the model. 
Through the seesaw mechanism, the other two right-handed 
 neutrinos play the role of realizing the observed 
 neutrino oscillation data. 
In this concise setup, 
 we have calculated the relic abundance of the dark matter. 
It has been found that the observed abundance can be reproduced 
 through the annihilation processes of 
 the dark matter particles enhanced by 
 resonances of either the SM Higgs boson, $B-L$ Higgs boson 
 or Z' gauge boson. 
As a result, the dark matter masses are almost fixed 
 to be half of the mass of either resonant states.

The classically conformal invariance and the $B-L$ symmetry 
 breaking via the CW mechanism 
 lead to a relation between model parameters, 
 namely, Higgs boson masses, Z' boson mass  
 and the dark matter mass are all related. 
We have identified the mass region to be consistent 
 with the experimental constraints. 
We have also calculated the spin-independent cross section 
 of the dark matter elastic scattering off with nuclei. 
The resultant cross section is found to be consistent 
 with the current limit by the direct dark matter detection 
 experiments. 
The parameter region can be tested in part 
 by the future experiments.

\section*{Acknowledgments}
We would like to thank Satoshi Iso 
 for useful discussions and comments. 
Y.O. would like to thank Department of Physics and Astronomy, 
 University of Alabama for hospitality during his visit.
The work of N.O. is supported in part 
 by the DOE Grants, No. DE-FG02-10ER41714.



\begin{thebibliography}{99}

\bibitem{seesaw}
P.~Minkowski, Phys. Lett. B {\bf 67}, 421 (1977);
T.~Yanagida, in \emph{Proceedings of the Workshop on the Unified
  Theory and the Baryon Number in the Universe} (O.~Sawada and
  A.~Sugamoto, eds.), KEK, Tsukuba, Japan, 1979, p.~95;
M.~Gell-Mann, P.~Ramond, and R.~Slansky, \emph{Supergravity} (P.~van
  Nieuwenhuizen et al. eds.), North Holland, Amsterdam, 1979, p.~315;
S.~L. Glashow, \emph{The future of elementary particle physics}, in
  \emph{Proceedings of the 1979 Carg{\`e}se Summer Institute
 on Quarks and Leptons} (M.~Levy et al. eds.),
 Plenum Press, New York, 1980, p.~687;
R.~N. Mohapatra and G.~Senjanovic,
 Phys. Rev. Lett. {\bf 44}, 912 (1980).



\bibitem{BL-LHC}
W.~Emam and S.~Khalil,
  Eur.\ Phys.\ J.\  C {\bf 522}, 625 (2007)
  [arXiv:0704.1395 [hep-ph]]; 
%
M.~Abbas and S.~Khalil,
  JHEP {\bf 0804}, 056 (2008)
  [arXiv:0707.0841 [hep-ph]].
%
M.~Abbas, W.~Emam, S.~Khalil and M.~Shalaby,
  Int.\ J.\ Mod.\ Phys.\  A {\bf 22}, 5889 (2008); 
%
K.~Huitu, S.~Khalil, H.~Okada and S.~K.~Rai,
  Phys.\ Rev.\ Lett.\  {\bf 101}, 181802 (2008)
  [arXiv:0803.2799 [hep-ph]]; 
%
 L.~Basso, A.~Belyaev, S.~Moretti and C.~H.~Shepherd-Themistocleous,
  Phys.\ Rev.\ D {\bf 80}, 055030 (2009)  
  [arXiv:0812.4313 [hep-ph]]; 
%
 L.~Basso, A.~Belyaev, S.~Moretti, G.~M.~Pruna and 
 C.~H.~Shepherd-Themistocleous,
  Eur.\ Phys.\ J.\ C {\bf 71}, 1613 (2011)  
  [arXiv:1002.3586 [hep-ph]].  



\bibitem{IOO1}
S.~Iso, N.~Okada and Y.~Orikasa,
 Phys.\ Lett.\ B {\bf 676}, 81 (2009)  [arXiv:0902.4050 [hep-ph]].


\bibitem{IOO2}
S.~Iso, N.~Okada and Y.~Orikasa,
 Phys.\ Rev.\ D {\bf 80}, 115007 (2009)  [arXiv:0909.0128 [hep-ph]].


\bibitem{IOO3}
S.~Iso, N.~Okada and Y.~Orikasa,
  Phys.\ Rev.\ D {\bf 83}, 093011 (2011)  [arXiv:1011.4769 [hep-ph]].  


\bibitem{RLG}
M.~Flanz, E.~A.~Paschos, U.~Sarkar and J.~Weiss,
  Phys.\ Lett.\  B {\bf 389}, 693 (1996)
  [arXiv:hep-ph/9607310]; 
%
A.~Pilaftsis,
  Phys.\ Rev.\  D {\bf 56}, 5431 (1997)
  [arXiv:hep-ph/9707235]; 
%
A.~Pilaftsis and T.~E.~J.~Underwood,
  Nucl.\ Phys.\  B {\bf 692}, 303 (2004)
  [arXiv:hep-ph/0309342].


\bibitem{OS}
 N.~Okada and O.~Seto,
 Phys.\ Rev.\ D {\bf 82}, 023507 (2010)  [arXiv:1002.2525 [hep-ph]].



\bibitem{vBL}
M.~S.~Carena, A.~Daleo, B.~A.~Dobrescu and T.~M.~P.~Tait,
  Phys.\ Rev.\  D {\bf 70}, 093009 (2004)
  [arXiv:hep-ph/0408098];
%
G.~Cacciapaglia, C.~Csaki, G.~Marandella and A.~Strumia,
  Phys.\ Rev.\  D {\bf 74}, 033011 (2006)  
  [arXiv:hep-ph/0604111].


\bibitem{LHCZ'} 
CMS-PAS-EXO-11-019



\bibitem{Basso} 
L.~Basso,
 arXiv:1106.4462 [hep-ph].


\bibitem{CW}
  S.~R.~Coleman, E.~J.~Weinberg,
  Phys.\ Rev.\  {\bf D7}, 1888-1910 (1973).


\bibitem{LHCreach}
  L.~Basso, A.~Belyaev, S.~Moretti and G.~M.~Pruna,
  JHEP {\bf 0910}, 006 (2009)  [arXiv:0903.4777 [hep-ph]].  


\bibitem{WMAP}
  D.~Larson, J.~Dunkley, G.~Hinshaw, E.~Komatsu, M.~R.~Nolta, C.~L.~Bennett, B.~Gold, M.~Halpern {\it et al.},
  Astrophys.\ J.\ Suppl.\  {\bf 192}, 16 (2011) 
  [arXiv:1001.4635 [astro-ph.CO]].


\bibitem{QCD sim1}
  H.~Ohki, S.~Aoki, H.~Fukaya, S.~Hashimoto, T.~Kaneko, H.~Matsufuru, J.~Noaki, T.~Onogi {\it et al.},
  PoS {\bf LAT2009}, 124 (2009).
  [arXiv:0910.3271 [hep-lat]].


\bibitem{QCD sim2}
  R.~J.~Crewther,
  Phys.\ Rev.\ Lett.\  {\bf 28}, 1421 (1972); 
  M.~S.~Chanowitz, J.~R.~Ellis,
  Phys.\ Lett.\  {\bf B40}, 397 (1972); 
  M.~S.~Chanowitz, J.~R.~Ellis,
  Phys.\ Rev.\  {\bf D7}, 2490-2506 (1973); 
  J.~C.~Collins, A.~Duncan, S.~D.~Joglekar,
  Phys.\ Rev.\  {\bf D16}, 438-449 (1977); 
  M.~A.~Shifman, A.~I.~Vainshtein, V.~I.~Zakharov,
  Phys.\ Lett.\  {\bf B78}, 443 (1978). 


\bibitem{XENON100} 
  E.~Aprile {\it et al.}  [XENON100 Collaboration],
  Phys.\ Rev.\ Lett.\  {\bf 107}, 131302 (2011)  
  [arXiv:1104.2549 [astro-ph.CO]].


\bibitem{XENON1T}
 The XENON Dark Matter Project
 \verb|http://xenon.astro.columbia.edu/XENON100_Experiment/|


\end{thebibliography}
\end{document}